# Ontology-Supported and Ontology-Driven Conceptual Navigation on the World Wide Web

*Michel Crampes and Sylvie Ranwez*

Laboratoire de Génie Informatique et d'Ingénierie de Production (LGI2P)
EERIE-EMA, Parc Scientifique Georges Besse
30 000 NIMES (FRANCE)
Tel: (33) 4 66 38 7000
E-mail: {Michel.Crampes, Sylvie Ranwez}@site-eerie.ema.fr

**ABSTRACT**
This paper presents the principles of ontology-supported and ontology-driven conceptual navigation. Conceptual navigation realizes the independence between resources and links to facilitate interoperability and reusability. An engine builds dynamic links, assembles resources under an argumentative scheme and allows optimization with a possible constraint, such as the user's available time. Among several strategies, two are discussed in detail with examples of applications. On the one hand, conceptual specifications for linking and assembling are embedded in the resource meta-description with the support of the ontology of the domain to facilitate meta-communication. Resources are like agents looking for conceptual acquaintances with intention. On the other hand, the domain ontology and an argumentative ontology drive the linking and assembling strategies.

**KEYWORDS :** Adaptive hypertext, conceptual navigation, metadata, XML, ontology, narration, time optimization, WWW.

**INTRODUCTION**
Navigation on the World Wide Web relies on two main technical basis, hypertext and Information Retrieval based on search engines. Each technique constitutes a paradigm, i.e. a specific way of looking at this World.

For the hypertext paradigm, the World Wide Web is a network of links between and within documents through which the user navigates using visual invitations (marks) on the documents. Meanwhile, IR search engines use key words and index databases to gather everything that may resemble a user's query. Each approach is very powerful and has proven to be efficient within its own paradigm. Practically, readers combine both. The lexical search is to look for unknown documents on specific topics, and the hypertext approach uses authors' links to complete the coverage of the topic as needed.

It is accepted that there is no ideal solution to a complex problem and a coherent paradigm may present limits when considering the complexity and the variety of the users' needs. Let's recall some of the traditional criticisms about hypertext. The readers get lost in hyperspace. The links are predefined by the authors and the author's intention does not necessary match the readers' intentions. There may be other interesting links to other resources that are not given. The narrative construction which is the result of link following may present argumentative pitfalls.

As regards the IR paradigm, there are other criticisms. The search engines leave the readers with a list of weighted documents having no other relation than the lexical one. The set of documents is a set of local results and there is no means for managing redundancy, or a lack of information. The order of presentation is often the decreasing order of the weights and there is no narrative construction between documents.

Beyond these specific criticisms, both approaches present other common limits. The reader is the one who must decide most of the navigation strategy. This responsability would not be a problem if the readers already knew the content of the documents they are invited to visit. But when the readers have very little idea about the documents, their content and their volume, which is usually the case, they have not enough information to decide what the best strategy is for meeting their goals.

Finally, no constraint is handled by the hypertext navigation on the behalf of the users, such as the time they have available to read the documents they access. This consideration has not inspired much research, but practically, this is the sort of constraint that influences quite a lot the readers' strategies.

The research project of our team is to define a new approach where an agent uses ontologies to work on the behalf of readers to find relevant documents, select among them the most appropriate, organize them, and establish links between them with a possible argumentative construction. During the work, the agent takes into account readers' requirements and constraints, particularly the readers' content objectives and their available time constraints.

This paper presents the principles of ontology-supported and ontology-driven conceptual navigation. Several possible models of conceptual navigation strategies are introduced. We illustrate two of them with different applications. We analyse the architectural differences and the advantages and disadvantages they bring about. As a conclusion, we show that what is at stake is not only adaptivity to the users' needs, but also interoperability and reusability.

**ONTOLOGY AND CONCEPTUAL NAVIGATION : PRINCIPLES AND EXAMPLE**

**Principle**
The general principle of ontology-supported conceptual navigation is the following :

- The system takes charge of the user's profile involving objectives and constraints.
- It automatically builds intentional weighted semantic links between documents or parts of documents.
- It gives roles (affordances, pragmatics) to these links, taking into account the ontology of the domain and an ontology of argumentation.
- It chooses among these links which are the best according to a particular context and a particular reader's intention.
- It assembles the resources using the most appropriate narrative or pedagogic strategy amongst possible strategies. During this computation, it complies with the user's time constraint, or any other economical constraint.

This approach differs from hypertext because resources have no specific links, be these explicit as in most hypertext systems, or implicit as in [27] where a visual spatial hypertext is used to show relationships and linkages between documents. In conceptual navigation links between resources and their narrative or pedagogical roles are computed from their description in a formal conceptual language and may vary according to the situation. The selection of the most pertinent resources and their organization are not controlled by the user, but are the result of a computation that involves an ontology of the domain, and possibly an argumentative ontology. It differs also from IR as far as resource retrieval is more semantically grounded with the ontology of the domain. The system does not only deliver a list of semantically related resources, but also an order of consultation to comply with domain content and narrative/pedagogy structures.

Ontology-supported conceptual navigation may use several navigation strategies. We introduce the first strategy with the Karina project (participants to all cited projects are acknowledged at the end of the paper).

**Example of application in adaptive course building : the Karina project**
The objective of the Karina project is to dynamically build courses which are adapted to the needs of a particular learner. The courses are composed of pedagogical resources that are available on line. Karina's long range objective is to propose several conceptual navigation strategies, among which the system will choose the best adapted to the learner's needs. For the moment, only the backward conceptual navigation strategy has been implemented. It will be discussed later on. Besides these strategies, Karina still allows for navigation using the traditional methods, i.e. word indexation and hyperlinks. Three main phases in the conceptual navigation process can be distinguished in Karina. These phases are summarized below. The first two phases are discussed in detail in other sections since they are at the core of conceptual navigation.

*Phase one: document selection and indexation.* The first phase is the production or the selection of resources that may be used or reused in the construction of training courses. These resources may have been produced either by a unique author or by different authors. Karina does not speculate on who is in charge of producing/selecting resources or how. The resources are indexed. A DTD (Document Type Definition), written in XML, is used to structure indexing. Help is obtained from indexing tools which propose a vocabulary and semantic constraints derived from an ontology of the domain.

*Phase two: Dynamic adaptive course building.* In order to build courses, Karina needs to know the learner's profile, i.e. the present knowledge, the knowledge objective and the learner's constraints. The main constraint which is considered is time. An engine called Conceptual Evocative Engine is in charge of selecting among the available indexed resources those that can entirely, or most often partly, fulfill the conceptual description of the learner's objectives. When chosen pedagogical material has prerequisites, those prerequisites become an intermediate objective for the engine (backward conceptual navigation). The result is a list of pedagogical resources which is ordered according to the objective-prerequisite navigation process.

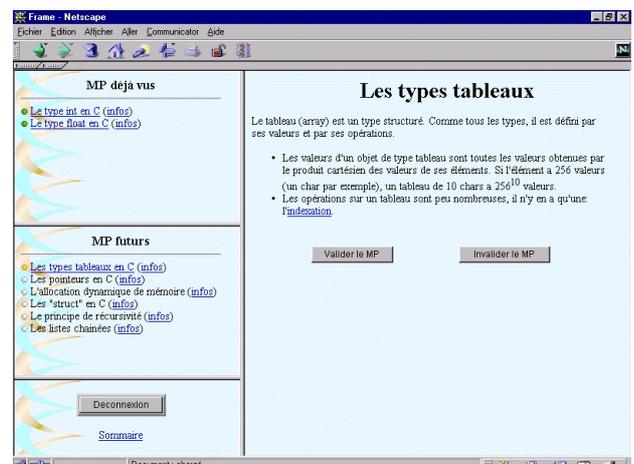

**Figure 1**

*Phase three: Courseware material presentation.* The titles of the selected pedagogical resources are presented in a small frame, down the left of the screen (see Figure 1). It is important to note that the resources are presented in the order of the conceptual backward navigation result. The resource is displayed on the right-hand side of the screen. Every time a resource is closed by the learner, it is considered to have been consulted and its title is erased from the bottom-left frame and rewritten in the top-left frame. All the resources in this frame can be consulted again at any time. For those resources that have not yet been consulted, a little dot on the left indicates with a color whether they are ready to be consulted or not depending on whether the prerequisites have been satisfied.

## ONTOLOGY-SUPPORTED CONCEPTUAL INDEXATION IN XML

### The Karina's DTD

The Karina's DTD[1] is a XML-written document which allows the qualification of complete resources, or parts of resources called "segments". The DTD is composed of several "elements" which contain most of the necessary information for retrieving a resource on a conceptual and argumentative basis, analysing it and assembling it with other resources [9]. In the following description of the DTD, we only discuss some features that are used for ontology-supported conceptual navigation, and more precisely for conceptual backward navigation :

<!ELEMENT description_conceptuelle (phrase_kldp+)> contains a conceptual description of the resource,
<!ELEMENT pre_requis (phrase_kldp+)> contains the conceptual prerequisites of the resource,
<!ELEMENT ontologie (%uri;) > contains the address of the ontology which is used for indexing,
<!ELEMENT temps_utilisation (#PCDATA)> contains the time value a learner should spend on the resource.

The time value is estimated by the tutor who indexes the resource. In future developments, this time value should be the result of statistical observations of readers' practices. The attribute "phrase_kldp" refers to Karina's Conceptual Language (KCL), which is described below.

### Karina's Conceptual Language (KCL)

This language is defined in the Karina DTD using XML. It formalizes conceptual descriptions of content into a structure called a Conceptual State Vector (CSV) presented in [8]. A CSV is a weighted sum of conceptual assertions. Each assertion is represented by a conceptual graph (CG) [28].

*Simplified Conceptual Graphs in Karina.* Although Sowa's CGs are very useful to formalize knowledge, they present some drawbacks in the context of Karina. They are not simple to use for a non-specialist. They are not easy to implement in order to be manipulated. They do not take into account the new languages being developed to describe meta-data for the Internet, particularly XML and RDF. In order to solve these different problems, Karina recognizes a simple form of CG, and uses some of its properties to retain its power of expression. They contain a maximum of three conceptual terms, the source, the destination and a predicate (a relation between the source and the destination). In order to deal with a complex knowledge content, it is necessary to use several of these graphs. For instance, the sentence (from 'Alice in Wonderland') "Alice looks at the Caterpillar for some time in silence" would be described in KCL with a conceptual graph $CG_i$ composed of several three term conceptual graphs :

$CG_i$ : { $CG_{i,1}$: [LITTLE_GIRL : Alice] → [LOOK_AT] → [CATERPILLAR : #],
$CG_{i,2}$ : [LITTLE_GIRL : Alice] → [LOOK_AT] → [SILENTLY],
$CG_{i,3}$ : [LITTLE_GIRL : Alice] → [LOOK_AT] → [TIME-PERIOD : #] }

The traditional CG relations like (AGNT) or (OBJ) have disappeared, but they are still implicit taking into account the ontology of the domain as it is explained below. As far as these three simple graphs describe the same situation, they can be merged applying Sowa's operation "copy", "restrict", "join", and "simplify" in order to rebuild the initial conceptual graph. To give more details to the situation, we simply need to add new assertions in the set. For example, if we want to enrich the $CG_i$ with the assertion that the caterpillar also speaks to Alice, we can add the following conceptual graph :

$CG_{i,4}$ : [ CATERPILLAR : #]] → [SPEAK_TO] → [LITTLE_GIRL : Alice }

*Conceptual typing with the help of the ontology of the domain.* An ontology is "an axiomatic characterization of the meaning of a logical vocabulary" [16]. It is modelled as a hierarchy of types and a set of relations beween those concepts which specify which assertions it is possible to make about a world corresponding to the domain. In Karina, semantic correctness and interoperability is supported by an ontology of the domain which is written in KCL. An ontology is stored as a resource specified with a particular DTD written in XML[1].

Karina's indexing interface makes use of the ontology of the domain to facilitate the indexing process and to prevent any mistakes. It opens up three slots for each Karina conceptual graph to be edited. The slots are constrained according to the ontology used for indexing the document. The first slot stands for the "source" of the conceptual graph. It contains the hierarchy of concepts from the ontology. When a concept is chosen, the indexer limits the hierarchical menu in the second slot to the concepts that are related to the source in the set of predicates in the ontology. It is then possible to choose in the concept hierarchy a concept among the children of these selected concepts. The same process is applied to the slot destination, with a hierarchical menu limited to the concepts that are in relation to the source and the predicate.

---
[1] The Karina DTD and the ontology DTD can be freely downloaded at the address:
http:// www.site-eerie.ema.fr/~multimedia

*Compatibility with RDF.* We have discussed the interest of this approach as far as simplicity and implementation are concerned. It is easier to implement a frame fixed in size in a database. This representation is also interesting because of its proximity to the RDF syntax which is the 'W3C Recommendation' for representing the semantics of resources on the Web [31]. In RDF, a "statement" is a "specific resource together with a named property plus the value of that property for that resource. These three individual parts of a statement are called, respectively, the subject, the predicate, and the object". Karina's {source, predicate, destination} triplet is very near the RDF's {subject, predicate, object} statement structure. Although when designing Karina we wanted to pay attention to a possible compatibility with RDF, we built our representation in a more practical and straightforward formalism. 'Karina Conceptual Language' lies between Sowa's conceptual graphs, and the RDF syntax, with particular attention paid to practicality.

**Conceptual State Vectors**
In order to emphasize specific statements, or concepts inside statements, each statement in the set of statements describing a resource is endowed with a weight having a real value between 0 and 1. A justification for this weight has been given in [7]. As a result, a Karina conceptual description of a resource is a Conceptual State Vector (CSV), i.e. a symbolic sum of weighted conceptual graphs.

**Translation and independant saving**
All the information entered for qualifying a resource is translated automatically into XML using Karina's DTD. It is a Resource Description (RD) which is stored in an independant file from the resource in order to avoid polluting a possible original meta-description of the resource. This choice is the result of several considerations :

- A resource can keep its genuine meta-description which has a specific meaning in the original context.
- The argumentative points of view may vary according to different tutors and there should be different RDs according to the different contexts.
- By keeping the resource in its original state, we partly avoid some problems with rights.
- Finally, it is easier to scan a separate meta-description stored in a database and it takes less space to store it. The meta-description can be local, and the resources distant.

**DEFINITION OF A LEARNER'S PROFILE**
In the present version of Karina, the learner's profile consists in three main specifications :

*The conceptual description of the learner's initial knowledge.* This description is done using a Conceptual State Vector which contains the main concepts and relations supposedly known by the learner. It is entered by the learners with a simplified interface to avoid manipulating conceptual graphs (see below).

*The conceptual description of the knowledge to be taught.* The knowledge objective to be taught is also described using a Conceptual State Vector. As for the learners' initial knowledge, since the CSV description is too formal for the learners to specify their goals, they are, in fact, invited to choose a set of sentences in natural language in a pre-specified list. Each sentence hides a CSV. To further simplify this approach, it is possible for the tutors to define conceptual curricula which consist of a list of contents and give it a name. It should be noted that the result of the specification of a content for Karina is not a list of pedagogical material, but a list of conceptual content descriptions.

*A time constraint.* The learner can give a time constraint to indicate how long he or she can spend on learning. This information is important. It allows Karina's engine to optimize the training and to avoid combinatorial explosion.

**BACKWARD CONCEPTUAL NAVIGATION STRATEGY**
When asked to design a course for a particular learner, Karina takes into account the initial profile and tries to build an ordered list of resources that can together meet the learner's objectives. To do so, it only considers the RDs that are available in the RD database. The resources may be stored locally or on another server. Their localisation will be taken into account only when they are presented to the learner. The backward conceptual navigation process is conducted in several steps. In each step, the process is not straightforward and several heuristics may be used on which we shall not go into detail. Karina's engine is written in Java and before analysis by the engine, the RDs are translated for manipulation into an Oracle relational database.

**The backward conceptual algorithm**

*Objective update.* The first step consists in updating the objective. Karina takes the CSV corresponding to the objective and withdraws those CGs that are present in the learner's initial model. The weights are not taken into account at this stage. This suppression is made with a total match between the slots of the CGs, i.e. when a slot is empty in one CG, and the corresponding slot is not empty in the other CG, the two CGs are considered not to match.

*Conceptual Proximity computation.* In a second step, the engine explores the different RDs and computes a match beween the learner's updated content objective and the conceptual contents of the resources. This process uses a unification algorithm to compute a Conceptual Proximity (CP) between two CSVs. This algorithm has been formally described in [7].

*Choice of the best resource.* The resource with the highest CP as regards the updated objective is selected. If several resources have the same CP value, Karina selects the one with the lower time value. This choice is justified because the shorter the resource, the more it will be possible to confine the course in the time constraint given by the learner. If two resources have the same duration, one is arbitrarily chosen. The other one is memorized in case the selection needs to be reviewed at the end (backtracking).

*Objective and profile updating.* Then Karina withdraws the content of the selected resource from the objective and adds this content to the learner's profile. It behaves as if the learner had consulted the resource. It also adds the prerequisites of the resource to the objective. When doing this, it only adds the prerequisites that are not already present in the learner's profile to avoid looking for contents that have already been dealt with by other selected resources or by the learner's initial knowledge. Any selected resource is tagged so that it will not be considered again during the following round of selection.

*End of selection.* The selection process ends when there is no content left in the objective, or if there are no resources matching the objective. The different resource durations are added up. If the result exceeds the learner's time constraint, Karina tracks back to choose the second-best selected resource in the queue which presents a shorter time value to try another path. If there is no path meeting the time constraint, Karina proposes the shortest path.

**Organisation of the resources**
The selected resources are finally presented in the order imposed by the prerequisites with the first on the list being the one requiring no other prerequisites than the learner's initial knowledge.

**OTHER CONCEPTUAL NAVIGATION STRATEGIES**
Traditional navigation strategies are still possible since the resources keep their original hyperlinks and the DTD allows the introduction of keywords for IR engines. But what is most interesting is the numerous conceptual navigation strategies that are possible. We present here some of them that we are studying and that are representative of the power of ontology-supported conceptual navigation.

**Conceptual expansion**
The backward navigation strategy presented above takes into account only the intitial goals and the different resource conceptual prerequisites to look for resources. Although this approach may lead to a good coverage of the domain being taught, it may leave aside interesting related concepts that are neither in the objectives nor in the prerequisites. In order to enlarge the conceptual domain of search, Karina's DTD provides a means for expanding the objectives with the element <!ELEMENT relation_conceptuelle (phrase_kldp+)>.

This element contains a conceptual description of other concepts or conceptual relations with which the content of the resource is in relation. Conceptual expansion consists in looking for resources that best match this conceptual description. It is a sort of open conceptual digression, a conceptual counterpart of links in hypertext.

Conceptual expansion may be applied in two ways. In the first case, the user may ask "more" about a subject when studying a resource, and the evocative engine will look for conceptually related resources. Since this conceptual relation may be attached to several segments of a resource, the expansion process may help to look in detail at different aspects of the content. The second type of conceptual expansion may be used by the application itself when there is a lack of material to build a sufficient delivery within the time constraint. In such a resource starvation context, the conceptual expansion policy allows for the filling up of the gaps. Conceptual expansion opens up many interesting possibilities that we are studying for other multimedia applications.

**Forward conceptual navigation**
Conceptual expansion can be used as a whole strategy which replaces backward navigation. A first resource is chosen and through conceptual expansion other resources are selected. In their turn, they may be used for expansion up to the point where the time constraint is reached. This process looks very much like free navigation in a hypertext, with the difference that here it is based on conceptual evocation and not hyperlinks. The risk is to get lost in a set of resources which are not linked through narrative constraints. It needs some conceptual railing.

**The conceptual specification strategy and its application in narrative abstraction**
The conceptual prerequisites and the conceptual relation constitute conceptual specifications for linking a resource to other resources. The advantage of embedding conceptual navigation specification within the resources is that the resources are independant, self-contained, and also cooperative. It is a first step to seeing resources as cooperative agents. The drawback is that the narrative/pedagogic strategy cannot be specified independantly from the resources. This drawback can be overcome with a strategy which is based on a conceptual specification of the expected final resource.

It consists in building a purely conceptual resource, i. e. an empty resource that only contains conceptual descriptions of segments. The engine goes to the first segment, takes its description as conceptual objectives and looks for resources that match these objectives. Then the engine proceeds to the next segment keeping the time constraint as a parameter for optimization. We have already presented this type of strategy implemented in the Godart project [8] which builds narrative abstraction from a linear narrative. If the application is educational, the conceptual content of a segment must be added to the learner's profile before going on to the next segment in order to avoid as much redundancy as possible. This strategy allows the formal specification of a conceptual pedagogical or narrative path, step by step, without requiring any resources to materialize it. When it is applied in different pedagogical contexts, i.e. with different pedagogical resource environments, it leads to different concrete pedagogical solutions.

## FROM ONTOLOGY-SUPPORTED TO ONTOLOGY-DRIVEN CONCEPTUAL NAVIGATION

### The domain ontology can drive the conceptual navigation

In the different conceptual navigation strategies described above, navigation has only been "supported" by the ontology of the domain, i.e. the ontology of the domain has been used only for indexing in order to be sure that the resources and the users' objectives are described with the same concepts and relations. Since an ontology of a domain contains a great deal of information about the domain, it is not surprising that we could use it in a more profitable way. [15] defines an "ontology-driven IS (Information System)" as an IS application where "the ontology is just another component (typically local to the IS), cooperating at run time towards the "higher" overall IS goal". Our central idea for ontology-driven conceptual navigation is to design an architecture where the engine only relies on ontologies for selecting resources, ordering them and adding a narrative/pedagogic intention during the linking process.

In pedagogic applications, this idea hinges on the observation that a table of content of a course looks very much like an ontology of the domain being taught. Titles and subtitles contain keywords that are presented in a hierarchy. Therefore, we can imagine that the ontology can be the basis for a training course when endowed with pedagogical properties. This is what we present in the next application example, Sybil.

### Example of an ontology-driven conceptual application

Sybil [25] teaches classical music such as the concept "sonata". Two ontologies have been formalized using conceptual graphs, the ontology of the domain and an ontology of pedagogy. As usual, the domain ontology contains a hierarchy of domain concepts and relations between the concepts. The pedagogic ontology contains a hierarchy of pedagogic concepts, pedagogic rules and pedagogic strategies. When given a learner's objective, the engine, written in the language Scheme, builds paths through resources which are conceptually described. A Resource Description is written in the conceptual graph formalism and contains the resource's content (it is limited to the concept the resource presents), its URL, its pedagogical role (i.e. exposure, example, explanation, test, etc. from the pedagogic ontology), and even its media (see Figure 2).

For example, when asked to explain what a sonata is, the engine browses the resources to find those that explain the concept. In order to select the most pertinent resources and to organize them in a proper pedagogic structure, the engine uses the resources' pedagogical roles from the RDs and the pedagogical rules from the pedagogic ontology. For instance, there is a rule which says: "IF an Explanation and an Example refer to the same topic, THEN the URL of the Explanation must precede the URL of the Example".

Moreover, if the general exposition strategy is "Top-Down", the engine will find in the domain ontology that a sonata is composed of four parts: the "exposition", the "development", the "recapitulation", and a "coda". These concepts become new goals for the exposition. As one can see, the conceptual navigation is driven by both the ontology of the domain and the pedagogic ontology, along with the RDs which contain the resources' conceptual description and pedagogic roles. The three structures are independant and reusable although there is a certain limit as far as the resources are concerned as we see next.

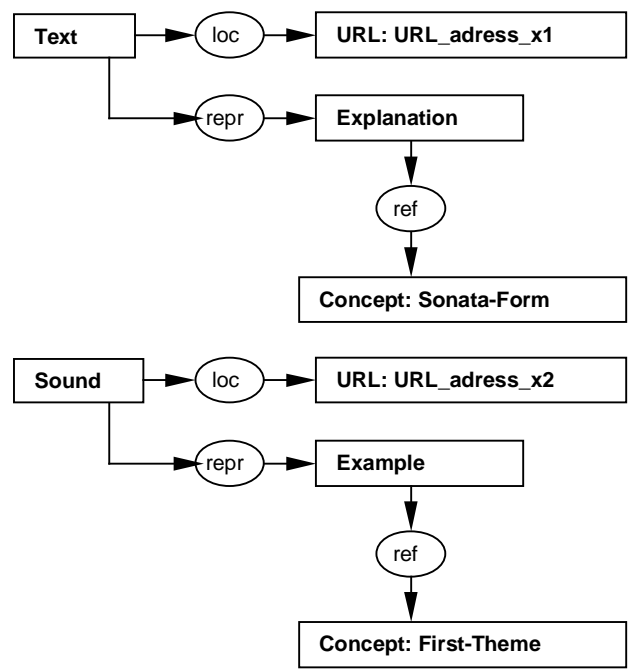

**Figure 2**

### Comparison of the two approaches

Both the Karina and the Sybil approaches are domain ontology-supported through indexation. In Karina, the conceptual navigation is the result of the engine strategies and the conceptual specifications embedded in the resources' description. In Sybil, the strategy is driven by the pedagogic ontology and the domain ontology. Both have pedagogic roles embedded in the resource descriptions. In Sybil, the pedagogic role is part of the resource description conceptual graph. In Karina, the element 'prerequisite' is a particular role for other related resources. There is also a specific element in the DTD called "type_pedagogique" which can be used to give a role to the resource.

The fact that the description of a resource contains the pedagogical role of the resource is very open to criticism because a resource may have several pedagogic roles according to the context. To solve this problem, we are working to have this role driven by the ontology, which means that it will be calculated through the ontology of the domain using the hierarchy property of concepts and relations, and the conceptual operations of the conceptual graph theory. Then the independence between the conceptual navigation strategies and the resources will be stronger, and all the material (ontologies, and resources) more interoperable and reusable.

**DISCUSSION AND RELATED WORKS**

Our research on ontology-supported/driven conceptual navigation stems from three main research origins, adaptive hypermedia, information retrieval and dynamic narrative constructions.

In adaptive hypermedia systems, the aim is to find a compromise between guiding users and letting them browse on their own [4,14,29,32]. These approaches are attempting to find ways of adapting pre-existent hypermedia. They do not aim at the construction of new links and their narrative organization in response to user needs is predefined.

The use of metadata to help with information retrieval and to share resources is a well-established practice. It is the basis of search engines such as Yahoo or Alta Vista when using indexes. But the efficacy of this brute force approach for computing similarities beween resources is limited by the biases caused by synonymy and polysemy (see [6] for a good insight into this problem). To avoid this pitfall, there are two possibilities.

The first one is to automatically build links under the constraint of an ontology which contains synonyms and relations between words (semantic networks). It is the case of Green [13] who automatically builds similarity links beween resources considering the fact that resources that are about the same thing will tend to use similar (although not necessary the same) words. He makes use of the WordNet database to build synset (sets of synonyms) weight vectors (the counterparts of Karina's conceptual state vectors).

The other possibility is to annotate resources under structural and semantical constraints to ensure interoperability [22]. Resource description articulates around complete resources, or parts of resources like in Karina, and makes use of either specific descriptors [2] or descriptors already established as standards or recommendations [11,21,17]. The XML (eXtensible Mark-up Language) [3] language allows the description of electronic resources by means of a DTD (Document Type Definition). The use of DTDs for describing Internet resources is a recent yet already well-established practice [19]. [1] proposes a DTD written in XML to describe the content of Audiovideo (AV) archives with meta-data. The authors also use an ontology to ascertain that several different resources are described with the same vocabulary. Then resource retrieval is based on dynamic linking either by taking an ontology or any resource as a point of entry. As far as only information retrieval is concerned, their approach is close to ours in many ways. We think, however, that the use of conceptual graphs and conceptual state vectors is more fruitful when it comes to building conceptual links. Moreover, our goal is also to build links with narrative commitment, and to comply with constraints, in particular the time constraint.

In [12], the authors propose an indexing approach based on conceptual graphs for information retrieval. They mention the support of an ontology and define a valuation function between two conceptual graphs in order to calculate their semantic proximity. Their objective is limited to information retrieval and they do not use any DTD to index the resources. When it comes to building coherent resources from different sources, information retrieval is not enough. We need a narrative/pedagogic structural organization. With the simple objective of building guided tours in hypermedia, Walden's Paths [26] allows teachers to define a directed path made up of resources extracted from the web out of their context, with annotations as meta-data in order to build a rhetoric structure. The introduction of narrative and pedagogic roles is one step towards the introduction of rhetorics. The roles of links and link typing have already been investigated [24,33]. But they are only considered as a means to add information to traditional links to help users build more consistent, argumentative paths.

In conceptual navigation, roles are dynamically defined in relation to the resource contents and the ontology of the domain. [5] investigates dynamic role construction. But conceptual navigation really begins when the systems dynamically assemble resources. Few examples can be found. [23] presents dynamic hypertext building based on NLG (Natural Language Generation) where the ontology of the domain clearly drives the composition along with linguistic rules. AHAM [10] comes close to our ontology-driven approach. It uses a domain model, a user model and a teaching model which consists of pedagogical rules to build adaptive hypermedia courses. Another similarity with Karina is the use of conceptual prerequisites. The AHAM authoring interface, still under study, will be based on the language Z, whilst Karina proposes an interface based on conceptual graphs to create XML compliant descriptions of resources. Sybil relies also on conceptual graphs to allow deep knowledge modelling and powerful conceptual operations.

In Karina's approach to conceptual navigation, the time constraint is used in order to prune the space search of related resources and to give a limit to the final delivery. This facility relies on the fact that the initial resources have been indexed with a time value which corresponds to the reading time hypothesized by the person who indexes. But, as [20] puts it, "reading time is a difficult thing to determine, and would benefit from further research". It may depend on the conceptual complexity of the content, its volume, the medium, the user's personality, intention, and level of acquaintance with the subject. Moreover, all these parameters may combine to define the reading time. Notwithstanding these difficulties, most teachers are able to give an average time length to the study of pedagogical material, be this value intuitive. We also explore a statistical definition of the time value from learners' practices. Other more objective constraints than time should also be taken into account, such as economical constraints when an on-line resource must be paid for. We think with Varian [30] that this temporal and economical dimension in hypermedia navigation should deserve more attention because it is clearly a very important parameter for efficient navigation on the World Wide Web.

## CONCLUSION: BEHAVIOR WITH (CONCEPTUAL) STRUCTURE

In [26], Rosenberg distinguishes two ways of approaching cybertext. On the one hand, hypertext systems are constructed with units bearing their own user interface, the navigation dynamics being localized in the units of text through visible links. On the other hand, full programmed cybertexts hide the dynamic hyper construction from the user. "The user is present as the algorithm unfolds its results, much as the viewer is present at the cinema". Rosenberg calls the first approach structural, and the second one behavioral. Conceptual navigation can be considered to be of the second kind. However, we do not expect an algorithm to build a meaningful path only through pure calculation even with good bricks of information. We have introduced two conceptual structures, the domain ontology and some argumentative ontology, be it narrative or pedagogic, to drive the hidden part of programmed cybertexts. Like any living entity, we think that programmed cybertexts should be endowed with a behavior and a structure. From such a point of view, the difference between hypertext and programmed cybertexts does no lie in the presence or the absence of a structure, but it is more about the nature of the structure, its localization, its visibility and which entity makes use of it. In hypertext, the structure is made up of visible links which are manipulated by the reader. In programmed cybertext, the structure is a deep hidden-knowledge structure which drives the engine. But most of all, in hypertext the structure is localized in the resources. In programmed cybertext, and particularly conceptual navigation, the structure is independant from the resources and the resources are self-contained. These last qualities allow better flexibility, interoperability and reusability of all components.

## ACKNOWLEDGMENTS
The Sybil project is sponsored by Digital Equipment, CEC Karlsruhe, Deutschland. The participants are Leidig T., from CEC Karlsruhe, Ranwez S. (main developper), and Crampes M., from Ecole des Mines d'Ales (EMA), France. Karina, was developped under a contract with the French Ministry of Industry. The developers are Crampes M., Bayard L., Gamrane N., Jacob O., Romagnoli E., from EMA, Gelly A., and Uny P., from Ecole Supérieure des Ingénieurs de Marseilles.